\documentclass[sigconf]{acmart}
\usepackage{soul}
\usepackage{graphicx}
\usepackage{booktabs}
\usepackage{float}
\usepackage{overpic}
\usepackage{float}  %设置图片浮动位置的宏包
\usepackage{subfloat}
\usepackage{stfloats} % for pic
\usepackage[skip=0.5\baselineskip]{caption}
\usepackage{bm}
\usepackage{amsmath}
\usepackage{booktabs}
\usepackage{multirow}
\usepackage{multicol}
\usepackage{enumitem}
\usepackage{subcaption}
\usepackage{threeparttable}
\usepackage{xspace}
\usepackage{balance}
\usepackage[normalem]{ulem}
\usepackage{algorithmicx,algorithm}
\usepackage{algpseudocode}
\usepackage{algorithm,algpseudocode}
\usepackage{marvosym}   % 添加通讯作者的信封
\usepackage{colortbl}   % 表格中的阴影颜色
\usepackage{tcolorbox}  % 文本加框
\tcbuselibrary{breakable}   % 配合文本加框，使得框可以跨页
\usepackage{wasysym}    % 添加实心圆空心圆
\usepackage[figuresright]{rotating} % 旋转表格
\usepackage{pifont} % 圆圈序号
\newcommand{\ie}{\emph{i.e.,}\xspace}
\newcommand{\eg}{\emph{e.g.,}\xspace}
\AtBeginDocument{%
	\providecommand\BibTeX{{%
			\normalfont B\kern-0.5em{\scshape i\kern-0.25em b}\kern-0.8em\TeX}}}

\copyrightyear{2026}
\acmYear{2026}
\setcopyright{cc}
\setcctype{by-nc-nd}
\acmConference[SIGIR '26]{Proceedings of the 49th International ACM SIGIR Conference on Research and Development in Information Retrieval}{July 20--24, 2026}{Melbourne, VIC, Australia}
\acmBooktitle{Proceedings of the 49th International ACM SIGIR Conference on Research and Development in Information Retrieval (SIGIR '26), July 20--24, 2026, Melbourne, VIC, Australia}
\acmDOI{10.1145/3805712.3809748}
\acmISBN{979-8-4007-2599-9/2026/07}
\begin{document}
\title{LLM-EDT: Large Language Models Enhanced Cross-domain Sequential Recommendation with Dual-phase Training}

\author{Ziwei Liu}
\authornote{Equal Contribution.}
\affiliation{%
  \institution{City University of Hong Kong}
  \city{Hong Kong}
  \country{China}
}
\email{lziwei2-c@my.cityu.edu.hk}

\author{Qidong Liu}
\authornotemark[1]
\affiliation{%
  \institution{Xi'an Jiaotong University \& City University of Hong Kong}
  \city{Xi'an}
  \country{China}
}
\email{qidongliu2-c@my.cityu.edu.hk}

\author{Wanyu Wang}
\affiliation{%
  \institution{City University of Hong Kong}
  \city{Hong Kong}
  % \state{Anhui}
  \country{China}
}
\email{wanyuwang4-c@my.cityu.edu.hk}

\author{Yejing Wang}
% \orcid{0000-0003-2852-9910}
\affiliation{%
  \institution{City University of Hong Kong}
  \city{Hong Kong}
  % \state{Guangdong}
  \country{China}
}
\email{yejing.wang@my.cityu.edu.hk}

\author{Pengyue Jia}
\affiliation{%
  \institution{City University of Hong Kong}
  \city{Hong Kong}
  % \state{Guangdong}
  \country{China}
}
\email{jia.pengyue@my.cityu.edu.hk}

\author{Tong Xu}
\affiliation{%
  \institution{University of Science and Technology of China}
  \city{He Fei}
  \country{China}
}
\email{tongxu@ustc.edu.cn}

\author{Wei Huang}
% \orcid{0009-0007-3641-7122}
\affiliation{%
  \institution{Independent Researcher}
  \city{Beijing}
  % \state{Guangdong}
  \country{China}
}
\email{hwdzyx@gmail.com}

\author{Chong Chen}
% \orcid{0000-0003-1417-2295}
\affiliation{%
  \institution{Tsinghua University}
  \city{Beijing}
  % \state{Guangdong}
  \country{China}
}
\email{cstchenc@163.com}
\author{Xiangyu Zhao}\authornote{Corresponding author}
% \orcid{0000-0003-2926-4416}
%\thanks{\Letter \ \text{Corresponding authors}}
\affiliation{%
  \institution{City University of Hong Kong}
  \city{Hong Kong}
  \country{China}
}
\email{xianzhao@cityu.edu.hk}

\renewcommand{\shortauthors}{Ziwei Liu, et al.}

\begin{abstract}
Cross-domain Sequential Recommendation (CDSR) has been proposed to enrich user-item interactions by incorporating information from various domains. Despite current progress, the domain imbalance issue and domain transition issue hinder further development of CDSR. The former presents a phenomenon where interactions in one domain dominate the entire behavior, leading to difficulty in capturing domain-specific features in the other domain. The latter points to the difficulty in capturing users' cross-domain preferences within the mixed interaction sequence, resulting in poor next-item prediction performance for specific domains. With world knowledge and powerful reasoning abilities, Large Language Models (LLMs) partially alleviate the above issues by functioning as both a generator and an encoder. However, current LLMs-enhanced CDSR methods are still under exploration, which fail to recognize the irrelevant noise and rough profiling problems. Thus, to address the aforementioned challenges, we propose an LLMs Enhanced Cross-domain Sequential Recommendation with Dual-phase Training (\textbf{LLM-EDT}). To address the domain imbalance issue while minimizing irrelevant noise, we propose the transferable item augmenter to adaptively generate possible cross-domain behaviors for users. Then, to alleviate the domain transition issue, we introduce a dual-phase training strategy to empower the domain-specific thread with a domain-shared background. As for the rough profiling problem, we devise a domain-aware profiling module to summarize the user's preference in each domain and adaptively aggregate them to generate comprehensive user profiles. The experiments on three public datasets validate the effectiveness of our proposed LLM-EDT. To ease reproducibility, we have released the detailed code online\footnote{https://github.com/Applied-Machine-Learning-Lab/SIGIR26\_LLM-EDT}.
\end{abstract}




\maketitle
\section{Introduction}\label{sec:intro}
\begin{figure}[!t]
	\centering
	\includegraphics[width = 0.95\linewidth]{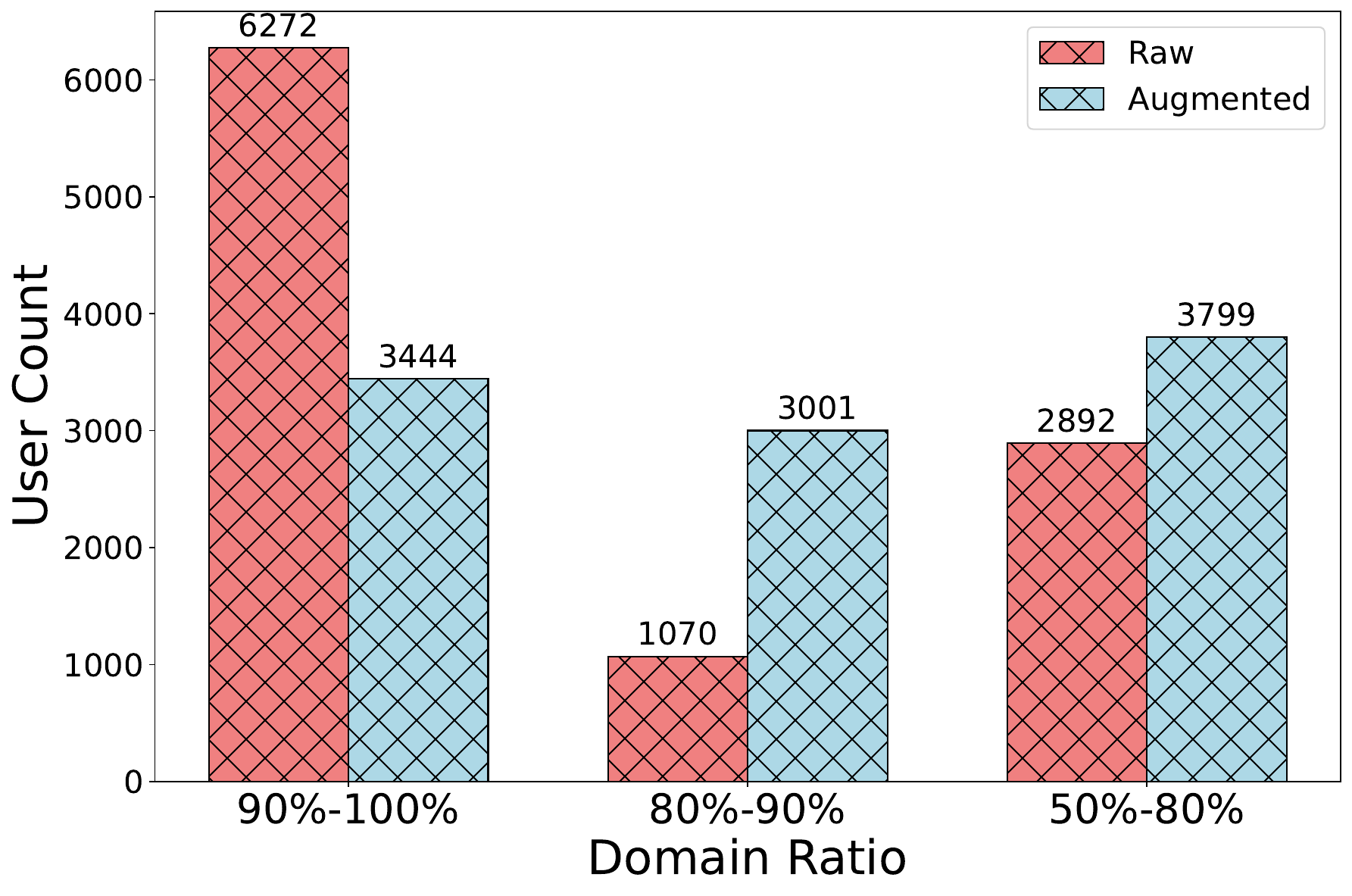}
	\caption{Illustration for imbalanced domain distribution. The X-axis represents the domain ratio, while the Y-axis represents the number of users under this ratio.}
        \label{fig:prel_exp}
\end{figure}
Sequential recommender systems (SRS), emerging as a technique to predict users' next interacted items, have empowered personalized services in different scenarios, such as streaming media~\cite{3} and e-commerce~\cite{1,2}. 
A surge of research has endeavored to leverage the capabilities of advanced deep learning architectures~\cite{zhang2025glint,wang2025star,liu2025sigma}. For example, GRU4Rec~\cite{GRU4Rec} leverages the recurrent neural networks (RNNs) to model different levels of sequence sessions, while SASRec~\cite{SASRec} utilizes the self-attention mechanism~\cite{attention} to capture users' preferences adaptively. Recent studies further explore bidirectional Transformer modeling~\cite{Bert4rec} and efficient alternatives such as pure MLP, multimodal MLP, sparse Transformer, and all-MLP architectures~\cite{li2022mlp4rec,li2023automlp,liang2023mmmlp,li2023strec,gao2024smlp4rec}. Researchers have also studied reinforcement learning and long-term optimization objectives in sequential recommendation~\cite{zhao2018deep,zhao2018recommendations,liu2024sequential}. However, the problem of data sparsity remains in recommendation~\cite{sparsity}, severely affecting the performance of SRS.

%As a promising approach, cross-domain sequential recommendation (CDSR) has been proposed to alleviate the data sparsity problem by learning user-item interactions from other domains in a mixed sequence (interaction with all domains)~\cite{CDSR1,CDSR2}. However, two challenges hinder the further development of CDSR.
As a promising approach, cross-domain sequential recommendation (CDSR) has been proposed to alleviate the data sparsity problem~\cite{sparsity}. By absorbing users' interactions from various domains to enrich the interaction sequence, CDSR can capture more comprehensive user preferences and give more accurate predictions. However, two issues hinder the further development of CDSR.
\begin{itemize}[leftmargin=*]
    \item \textbf{i) Domain Imbalance Issue}: %Imbalanced Domain Distribution}:  
    This issue stems from the imbalance of items from different domains in the user's interactions. For illustration, we define the domain ratio $r$ as the proportion of domain \textit{A} items among all the items in one user's interaction. For a two-domain situation, it can be calculated by $r =\frac{|n^A-n^B|}{n^A+n^B}$, where $n^A$ and $n^B$ respectively represent the number of items from domain \textit{A} and \textit{B} in one user's interaction histories. Then, we present the detailed statistics of Amazon dataset (Cloth - Sport)\footnote{The Cloth domain refers to domain A, while the Sport domain refers to domain B} in Figure~\ref{fig:prel_exp}. Observing from the results, the majority of users (red label) are located in the `90\%-100\%' category, indicating the severe domain imbalance in practice. Such an imbalance may lead to difficulty in capturing the domain-specific features, severely affecting the recommendation performance~\cite{cdsrsurvey1,bridge}. To alleviate this problem, some existing methods~\cite{transfer1,transfer2} employ transfer learning frameworks to enhance information flow across domains. Others~\cite{gnnbased} apply graph neural networks (GNNs)~\cite{gnn,gnnrec} to bridge the domain gap by modeling item connections. However, these methods heavily rely on overlapping users (those with interactions in both domains), which hinders the applications in real-world scenarios~\cite{cdsrsurvey2}. 
    %This challenge stems from the imbalanced distribution of domains within user-item interactions. As illustrated in Figure~\ref{fig:prel_exp}, the majority of users in the original dataset (red label) are located in the `Dominant' categories, indicating a severe imbalance in domain distribution. This concentration of user preferences within a single domain results in the neglect of latent cross-domain behavioral patterns in sequential recommendations.
    %To alleviate this problem, some existing methods~\cite{transfer1,transfer2} employ transfer learning frameworks incorporating multiple contrastive objectives to transfer information across domains. Others~\cite{gnnbased} apply graph neural networks (GNNs)~\cite{gnn} to bridge the domain gap by modeling item connections. However, these methods heavily rely on overlapping users (those with interactions in both domains), which limits their applicability in practical scenarios~\cite{cdsrsurvey2}. 
    \item \textbf{ii) Domain Transition Issue}: The other issue originates from the difficulty in capturing the user's cross-domain preference within the mixed interaction sequence. Current research adopt two paradigms to alleviate this problem. A line of them~\cite{tpuf_single,preprec_single} only leverage the global domain-shared features, but employ extra training objectives to identify the transition pattern between domains. Nevertheless, they lack the necessary details within domain-specific features, making it even more difficult to recognize the transition pattern. Others~\cite{CDSR1,CDSR_tri,TriCDR,ABXI} introduce extra domain-specific threads to alleviate this problem. By training the domain-specific threads and the domain-shared thread in parallel, they can easily recognize users' preferences in a specific domain. However, they fail to establish the necessary connection between domain-specific and domain-shared features, resulting in relatively ambiguous modeling for domain-specific features.
    %\item \textbf{ii) Insufficient Domain Representation}: The core challenge stems from the tension between global cross-domain patterns and local domain-specific features. While cross-domain sequences preserve comprehensive interaction patterns across domains, they lack fine-grained representations within individual domains. Conversely, single-domain sequences capture rich intra-domain features but fail to model inter-domain dependencies. To address this, methods such as TriCDR \cite{TriCDR} employ multi-threaded architectures to process global and local information concurrently. However, these approaches fail to effectively incorporate cross-domain contextual information when training the domain-specific thread, affecting the performance in individual domains.
\end{itemize}

Benefiting from the capability of Large Language Models (LLMs) in reasoning and representation~\cite{reasoning,representation,liu2025large}, the aforementioned issues can be partially alleviated. By leveraging LLMs to perform data augmentation and training a meta network to filter the noise, LLMCDSR~\cite{llmcdsr} alleviates the \textbf{Domain Imbalance Issue} as shown in the blue columns of Figure~\ref{fig:prel_exp} by utilizing a LLMs-based data augmentation. Nevertheless, it sacrifices the training and inference efficiency while still introducing irrelevant samples, leading to the \textbf{\ding{182} Irrelevant Noise} problem. For the \textbf{Domain Transition Issue}, another approach, LLM4CDSR~\cite{bridge}, employs LLMs as both the encoder and summarizer. However, it only enhances the item and user representations with LLMs, while still following the training paradigm of previous research~\cite{CDSR1,CDSR_tri,TriCDR}. This paradigm fails to learn the user preference from different granularities in a gradual manner, \ie, from global to domain-specific, which may hinder the user's preference updating for domain-specific items, further exacerbating the \textbf{Domain Transition Issue}. Moreover, naively introducing LLMs as profile summarizers raises the \textbf{\ding{183} Rough Profiling} problem. Such a way only leverages one-step summarization to generate the preference, which fails to capture the fine-grained features from each domain. 

To address these challenges while further exploring LLMs' capabilities in CDSR, we propose an innovative framework named \textbf{\underline{L}}arge \textbf{\underline{L}}anguage \textbf{\underline{M}}odels \textbf{\underline{E}}nhanced Cross-domain Sequential Recommendation with \textbf{\underline{D}}ual-phase \textbf{\underline{T}}raining (LLM-EDT). To alleviate the domain imbalance issue while mitigating the irrelevant noise that LLMs introduce, we devise a transferable item augmenter to enrich user-item interactions. To address the domain transition issue, we propose a dual-phase training framework that empowers domain-specific threads with a domain-shared background acquired by pre-training the model on a mixed sequence. In terms of the rough profiling, a domain-aware profiling module is introduced, which leverages a summarize-reform-analyze pipeline to generate comprehensive and fine-grained user profiles. Notably, since both the LLMs item embeddings and generated user embeddings can be cached in advance, our framework eliminates LLMs inference while serving, maintaining efficiency. The contribution of our proposed LLM-EDT can be concluded as follows: 
\begin{itemize}[leftmargin=*]
	\item To address the domain imbalance issue while mitigating the irrelevant noise, we propose a transferable items augmenter. To prevent LLMs from generating rough user profiles, we designed the domain-aware profiling.
	\item To alleviate the domain transition issue, we propose a dual-phase training framework to empower the domain-specific thread with well-trained information from the domain-shared thread.
    \item Comprehensive experiments on three public datasets validate the effectiveness of LLM-EDT.
\end{itemize}

\section{Preliminary}
\noindent \textbf{{Data Augmentation.}} Let $u \in \mathcal{U}=\left\{{u}_{1},\dots,{u}_{\left|\mathcal{U}\right|}\right\}$ denotes the user in the user set. For a two-domain situation in CDSR, the original item set $\mathcal{I}$ can then be separated into two sets (\textit{A} and \textit{B}), \ie $\mathcal{A}=\left\{a_1,a_2,\cdots,a_{\left|\mathcal{A}\right|}\right\}$ and $\mathcal{B}=\left\{b_1,b_2,\cdots,b_{\left|\mathcal{B}\right|}\right\}$. $\left|\mathcal{U}\right|$, $\left|\mathcal{A}\right|$ and $\left|\mathcal{B}\right|$ respectively represents the number of users, items of domain \textit{A} and items of domain \textit{B}. For a chronologically ordered interaction sequence in general SRS, the user sequence in mixed and individual domain can then be defined as ${S}_{u}=\left\{v_1^u,v_2^u,\dots,v_{n}^u\right\}$, ${S}_{u}^{A}=\left\{a_1^u,a_2^u,\dots,a_{n^A}^u\right\}$ and ${S}_{u}^{B}=\left\{b_1^u,b_2^u,\dots,b_{n^B}^u\right\}$, where $v_i$ belongs to the union set of $\mathcal{A}$ and $\mathcal{B}$ and $n =n^A+n^B$. $n^A$ and $n^B$ respectively represent the lengths of interaction sequences in domain \textit{A} and \textit{B} for a specific user $u$. Noted that we omit the mark $(u)$ in the following sections for simplicity. 
By leveraging LLMs, we conduct data augmentation (See in Section~\ref{sec:sample_generator}) to generate $k$ possible cross-domain behaviors, \ie $\left\{\hat{a}_{1},\hat{a}_{2},\dots,\hat{a}_{k}\right\}$ and $\left\{\hat{b}_{1},\hat{b}_{2},\dots,\hat{b}_{k}\right\}$. With our designed insertion mechanism mentioned in Section~\ref{sec:insertion}, we can insert the acquired samples into the mixed sequence $S$ and formulate the augmented sequence as:
\begin{equation}
\label{equ:aug_seq}
\begin{aligned}
    {S}_{aug}&=\left\{v_1,\hat{v}_2,\dots,v_{n+2k-1}, \hat{v}_{n+2k}\right\}\\
\end{aligned}
\end{equation} 
where $\hat{v}_i$ represents the inserted augmented interaction. Then, we can divide $S_{aug}$ into different domains chronologically as:
\begin{equation}
    \begin{aligned}
    {S}_{aug}^{A}&=\left\{a_1,\hat{a}_2,\dots,a_{n^A+k}\right\}\\
    {S}_{aug}^{B}&=\left\{b_1,\hat{b}_2,\dots,b_{n^B+k}\right\}
    \end{aligned}
\end{equation}

\vspace{1mm}
\noindent \textbf{Cross-domain Sequential Recommendation Objective.} 
Leveraging the aforementioned notations, CDSR task can be formulated by transforming the general SRS problem across two domains as:
\begin{equation}
\begin{aligned}
\arg\max_{v_i \in \mathcal{A}_{aug}} P(v_{n+1} &= v_i |{S}_{aug}, {S}_{aug}^{A}, {S}_{aug}^{B})\\
\arg\max_{v_i \in \mathcal{B}_{aug}} P(v_{n+1} &= v_i |{S}_{aug}, {S}_{aug}^{A}, {S}_{aug}^{B})
\end{aligned}
\end{equation}
\begin{figure*}[!t]
	\centering
	\includegraphics[width = \linewidth]{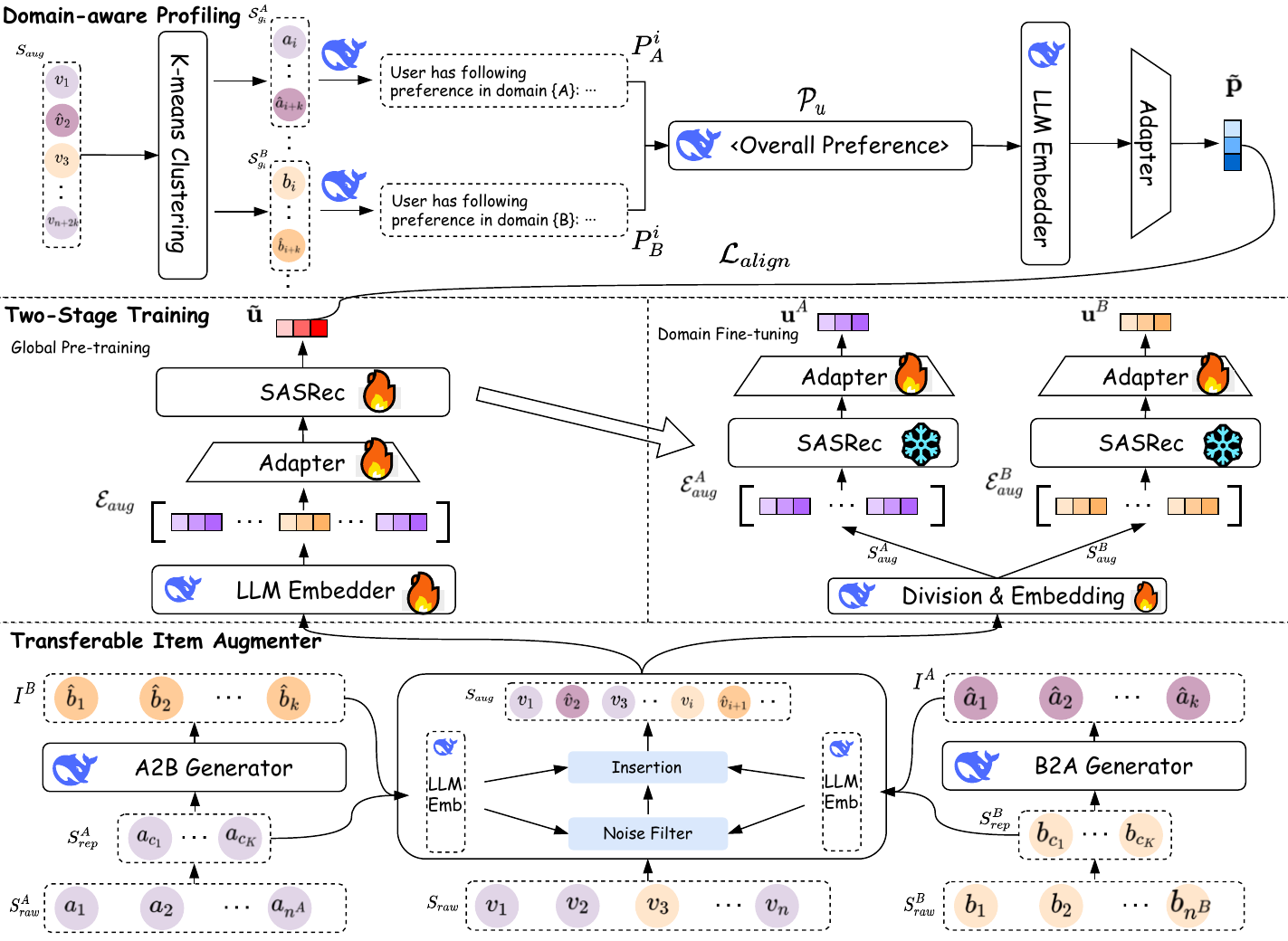}
	\caption{Framework of proposed LLM-EDT}
        \label{fig:Overview}
        \vspace{-4mm}
\end{figure*}

\section{Methodology}
This section delineates the architecture of the proposed LLM-EDT framework. We begin with an overview of the framework, followed by detailed descriptions of its three modules: transferable item augmenter, dual-phase training framework, and domain-aware profiling. Finally, we specify the training and inference procedures.

\subsection{Overview}
In this section, we present an overview of our proposed framework in Figure~\ref{fig:Overview}. To address the domain imbalance issue, we first propose the \textbf{Transferable Item Augmenter} to generate samples from one domain by learning from the other domain. Then, by leveraging the designed noise filter and insertion procedure, we can obtain balanced interaction sequences in each domain, \ie, ${S}_{aug}^{A}$ and ${S}_{aug}^{B}$. After that, a \textbf{Dual-phase Training Framework} is introduced to empower the domain-specific threads with a domain-shared background to address the domain transition issue. It first pre-trains the base model with mixed interaction embedding sequence $\mathcal{E}_{aug}$ in the stage of \textbf{Global Pre-training}, then gradually learns the individual embedding sequence, \ie $\mathcal{E}_{aug}^A$ and $\mathcal{E}_{aug}^B$, in the \textbf{Domain Fine-tuning} stage to derive the final next-item prediction for both domains. At last, we introduce \textbf{Domain-aware Profiling} to alleviate the problems naive LLM profiling causes. Using a clustering process, the mixed interaction sequence can be grouped into $S_{g_i}^A$ and $S_{g_i}^B$. Then, we perform a summarize-reform-analyze pipeline to generate fine-grained textual profiles for users. By leveraging LLMs as encoders, we can obtain the final user preference $\mathbf{\tilde{p}}$, which is then injected into the dual-phase training framework to construct the loss $\mathcal{L}_{align}$.

\subsection{Transferable Item Augmenter}
To address the domain imbalance issue, we would like to balance the domain ratio described in Section~\ref{sec:intro} by generating items for the "less dominant" domain. The derived items are transferable because they are relatively similar to the dominant domain. To achieve this, we propose a sample generator that leverages LLMs' semantic understanding and reasoning abilities. Recent foundation models such as DeepSeek-V3 and DeepSeek-R1 further demonstrate the practical potential of strong generation and reasoning capabilities in LLMs~\cite{deepseek,deepseekr1}. Then, to mitigate irrelevant noise, we introduce a noise filter to remove irrelevant samples. Finally, since position information is crucial for capturing the user's interest shift in CDSR, we devise an insertion module to serve as an effective bridge between domains.
%Currently, LLMs have emerged as a powerful technique to perform item-level augmentation~\cite{llmins,samplellm,llmcdsr}. Nevertheless, the application for the cross-domain situation is still under exploration. Some methods~\cite{llmins,samplellm} neglect the correlation between domains, while others~\cite{llmcdsr} introduce unavoidable irrelevant items and noises. Moreover, neither of them recognizes the importance of items that belong to one domain but have relatively high similarity with another domain, which we defined as transferable items. To mitigate these problems, we design the Transferable Item Augmenter, featuring a sample generator, a noise filter, and an insertion module. 
\subsubsection{\textbf{Sample Generator}}\label{sec:sample_generator}
Since our goal is to generate transferable items for users who suffer from the domain imbalance issue, we devise two LLMs-based generators to generate possible interacted items from two perspectives, \ie \textit{A2B} and \textit{B2A}.  Specifically, in the \textit{A2B} perspective, the items in domain \textit{A} will first be grouped using the K-means method~\cite{kmeans}, which ensures the diversity of the selected samples. Then, regarding these clusters as representative items, we can obtain the diverse samples, denoted as ${S}_{rep}^A$, from the original interaction sequence ${S}_{raw}^{A}$. Lastly, leveraging a well-constructed prompt $T$, we can acquire the generated transferable items in domain \textit{B}, denoted as $I^B$.
Similarly, we can also obtain the transferable items in domain \textit{A} via the same procedure. Thus, the whole generation process can be concluded as:
\begin{equation}
\begin{aligned}
     I^B &= \operatorname{LLMs}(T,S^A_{rep})\\
     I^A &= \operatorname{LLMs}(T,S^B_{rep}) 
\end{aligned}
\end{equation}  
%\begin{figure}[!t]
%	\centering
%	\includegraphics[width = \linewidth]{FigureMaking/item_prompt.pdf}
%	\caption{Item prompt template.}
%        \label{fig:prompt}
%\end{figure}
\subsubsection{\textbf{Noise Filter}}\label{sec:item_temp}
LLMs often suffer from the problem of hallucination, tending to generate noisy samples~\cite{hallucination}. 
To alleviate this problem, we design a novel filtering strategy to filter the irrelevant samples in the generated $I_B$ and $I_A$. 
Since LLMs embeddings are proven to capture the semantic relationships effectively~\cite{representation,SAID,llmretrieval}, we first employ LLMs as an encoder to generate item embeddings. Then, rather than leveraging meta-learning to recall the generated samples~\cite{llmcdsr}, we perform noise filtering to refine the augmentation results before training the full framework. In detail, for both the raw interactions and the generated samples in each domain, we first group the items' text attributes, \ie name, brand, category, and description, by utilizing a designed item prompt template. Then, we adopt LLMs text embedding model\footnote{\url{https://api.deepseek.com/}} to get the corresponding item representation $\mathcal{E}_{rep}^A$, $\mathcal{E}_{rep}^B$, $\mathcal{E}^{I^A}$ and $\mathcal{E}^{I^B}$. To filter the irrelevant items from the generated samples while maintaining the effective ones, we calculate two sets of cosine similarity scores for a specific item $i \in I^A \cup I^B$ and select the highest scores with the corresponding items as follows:
\begin{equation}
\label{equ:posi}
\begin{aligned}
    (s^A_{i}, score^A_{i}) &= \arg\max_{s_k^A \in S^A_{rep}} \operatorname{sim}(\mathcal{E}_i, \mathcal{E}_{s_k^A})\\
(s^B_{i}, score^B_{i}) &= \arg\max_{s_k^B \in S^B_{rep}} \operatorname{sim}(\mathcal{E}_i, \mathcal{E}_{s_k^B})
\end{aligned}
\end{equation}
%\begin{equation}
%\begin{aligned}
%    \mathcal{S}^A_i &= \{\operatorname{sim}(\mathcal{E}_i, \mathcal{E}_{s_k^A}) | s_k^A \in S^A_{rep}\}\\
%\mathcal{S}^B_i& = \{\operatorname{sim}(\mathcal{E}_i, \mathcal{E}_{s_k^B}) | s_k^B \in S^B_{rep}\}
%\end{aligned}
%\end{equation}
where $\mathcal{E}_i$, $\mathcal{E}_{s_k^A}$ and $\mathcal{E}_{s_k^B}$ respectively represent the embeddings of item $i$, ${s_k^A}$ and ${s_k^B}$. $s_i^A, s_i^B$ respectively represent the corresponding position for $score_i^A, score_i^B$. %After that, we select the highest scores with the corresponding items as follows:
For the \textit{A2B} perspective, as discussed previously, our goal is to generate the user's latent interest in domain \textit{B} by feeding the representative domain \textit{A} interactions into LLMs. Thus, for a generated item $i$ in $I^B$, the process can be formulated as follows:
\begin{equation}
\begin{aligned}
\label{equ:tau}
\mathcal{T}_{A2B} = \{i \in I^B &| \operatorname{rank}(score^A_{i}) \leq k \land score^B_{i} \geq \tau\}
\end{aligned}
\end{equation}
where $\mathcal{T}_{A2B}$ denotes the obtained samples after filtering and $\operatorname{rank(\cdot)}$ operation denotes a sorting and selection procedure. Notably, we can also perform the filtering in \textit{B2A} perspective via a similar procedure to acquire the $\mathcal{T}_{B2A}$.
\subsubsection{\textbf{Insertion}}\label{sec:insertion}
Research has proven that the insertion position is also vital for data augmentation~\cite{jiao2024rethinking}.
Unlike current augmentation methods~\cite{samplellm,llmcdsr} that randomly insert the generated items into the sequence, we strictly insert the filtered samples, \ie $\mathcal{T}_{A2B}$ and $\mathcal{T}_{B2A}$, according to the calculated positions in equation~\eqref{equ:posi}. For example, for a generated item $i$ of domain \textit{B} in $\mathcal{T}_{A2B}$, we acquire the most similar domain \textit{A} item $s_i^A$ in the raw sequence by leveraging equation~\eqref{equ:posi}. Since the transferable items should act as a signal for domain shifting, the generated sample of domain \textit{B} will be inserted right after the corresponding item of domain \textit{A}, \ie $s_i^A$, to construct a bridge between domain \textit{A} and \textit{B}. After performing similar procedures iteratively, we can insert all the items in $\mathcal{T}_{A2B}$ and $\mathcal{T}_{B2A}$ into the raw interaction and acquire $S_{aug}$ as mentioned in equation~\eqref{equ:aug_seq}.
%we can acquire the final augmented sequence, \ie $S_{aug}$, by inserting $i$ into the original all-domain sequence $S_{raw}$ according to the position of $s_i^A$ to further construct a bridge between domain \textit{A} and \textit{B}. 
\subsection{Dual-phase Training Framework}
To address the transition problem, some previous CDSR methods~\cite{bridge,TriCDR,CDSR1} adopt a three-thread framework to capture both domain-specific and domain-shared preferences. However, they usually organize the different threads in parallel, falling short of establishing an effective connection between them. To address this limitation, we propose a dual-phase training framework featuring global pre-training and domain fine-tuning. This training strategy leverages the background knowledge pre-trained by the domain-shared thread to empower domain-specific threads, further improving the cross-domain recommendation performance.

\subsubsection{\textbf{Global Pre-training}} 
In the pre-training stage, we train a base SASRec model~\cite{SASRec} on augmented mixed sequences $S_{aug}$ to endow it with global knowledge. Specifically, it begins with a LLM Embedder, which derives the LLMs embeddings of $S_{aug}$ by leveraging the same instruction in Section~\ref{sec:item_temp}, denoted as $\mathcal{E}_{aug}=\left\{\tilde{\mathbf{e}}_1,\tilde{\mathbf{e}}_2,\dots,\tilde{\mathbf{e}}_{n+2k}\right\}$, where $\tilde{\mathbf{e}}_i \in \mathbb{R}^{d_L}$ is the item embedding, and $d_L$ is the embedding size of LLMs. Then, following the design of current methods~\cite{LLMemb}, we leverage a trainable adapter to project LLMs semantic space to recommendation space while transforming the embedding size of LLMs $d_L$ to the size of the general SRS $d$. The whole process can be formulated as follows:

\begin{equation}
\label{equ:reducedi}
    \tilde{\mathbf{e}}_i = \mathbf{W}_1(\mathbf{W}_2\mathbf{e}_i + \mathbf{b}_2) +\mathbf{b}_1
\end{equation}
where $\mathbf{W}_1 \in \mathbb{R}^{d\times \frac{d_L}{2}}$, $\mathbf{W}_2 \in \mathbb{R}^{\frac{d_L}{2}\times d_L}$, $\mathbf{b}_1 \in \mathbb{R}^d$, $\mathbf{b}_2 \in \mathbb{R}^{\frac{d^L}{2}}$ are the parameters of adapter. After that, the SASRec~\cite{SASRec} backbone is imposed on $S_{aug}$ to capture the global preference, encoded into the global user representation $\tilde{u} \in \mathbb{R}^{d}$. For a given user $u$ and item $i$, the whole prediction process can then be formulated as:
\begin{equation}
\begin{aligned}
    \tilde{\mathbf{u}} &= f_{\tilde{\theta}}(S_{aug})\\
    P(v_{n+2k+1}&=v_i|S_{aug}, v_i \in \mathcal{I}_{aug})=\tilde{\mathbf{e}}_i^T\cdot\tilde{\mathbf{u}} 
\end{aligned}
\end{equation}
where $P(\cdot)$ represents the prediction probability that item $i$ will be preferred by user $u$. Then, we can formulate the loss function for the first pre-training stage as:
\begin{equation}
\begin{aligned}
      \mathcal{L}_{pre} = - \sum_{u \in \mathcal{U}}\sum_{j=1}^{n+2k}\log \sigma(P(v_{j+1}=v^+)-P(v_{j+1}=v^-))  
\end{aligned}
\end{equation}
where $v^+$ and $v^-$ respectively represent the ground truth and the randomly sampled negative item. 

\subsubsection{\textbf{Domain Fine-tuning}}
After empowering the base model with pre-trained domain-shared information, we then perform the domain fine-tuning process to extract the domain-specific preference for each domain. It begins with a division procedure. As previously discussed, the constructed $S_{aug}$ is chronologically ordered by leveraging our insertion strategy. We can directly base on the domain notation to extract two domain-specific sequences, \ie $S^A_{aug}$ and $S^B_{aug}$.  Then, we perform a similar LLMs embedding process to obtain the item embedding for domain \textit{A} and domain \textit{B}, denoted as $\mathcal{E}_{aug}^A=\left\{\tilde{\mathbf{e}}_1^A,\tilde{\mathbf{e}}_2^A,\dots,\tilde{\mathbf{e}}_{n^A+k}^A\right\}$ and $\mathcal{E}_{aug}^B=\left\{\tilde{\mathbf{e}}_1^B,\tilde{\mathbf{e}}_2^B,\dots,\tilde{\mathbf{e}}_{n^B+k}^B\right\}$. After that, the frozen SASRec backbone is imposed on $S^A_{aug}$ and $S^B_{aug}$, and two trainable adapters are introduced to encode into the corresponding user representation $\mathbf{u}^A\in \mathbb{R}^d$ and $ \mathbf{u}^B\in \mathbb{R}^d$. The whole process can be formulated as follows:
\begin{equation}
\begin{aligned}
    \mathbf{h}^A&=f_{\tilde{\theta_{F}}}(S^A_{aug}), \mathbf{h}^B=f_{\tilde{\theta_{F}}}(S^B_{aug})\\ 
    \mathbf{u}^A &= \mathbf{W}_3(\mathbf{W}_4\mathbf{h}^A + \mathbf{b}_4) +\mathbf{b}_3\\
    \mathbf{u}^B &= \mathbf{W}_5(\mathbf{W}_6\mathbf{h}^B + \mathbf{b}_6) +\mathbf{b}_5
\end{aligned}
\end{equation}
For a given user $u$ and item $i$ in domain \textit{A}, the prediction probability can be calculated by leveraging both the global representation $\tilde{\mathbf{u}}$ and domain-specific representation $\mathbf{u}^A$:
\begin{equation}
\label{equ:loss1}
P(v_{n^A+k+1} = v_i \mid S_{aug},S_{{aug}}^A, v_i \in \mathcal{A}_{{aug}}) 
= (\tilde{\mathbf{e}}^A_i)^{{T}} \cdot \mathbf{u}^A + \tilde{\mathbf{e}}_i^T\cdot\tilde{\mathbf{u}}
\end{equation}
Hence, the loss function for domain \textit{A} can be calculated as follows:
\begin{equation}
\label{equ:loss2}
\begin{aligned}
      \mathcal{L}_{A} = - \sum_{u \in \mathcal{U}}\sum_{j=1}^{n^A+k}\log \sigma(P(v_{j+1}=v^+)-P(v_{j+1}=v^-))  
\end{aligned}
\end{equation}
where $v^+$ and $v^-$ respectively represent the ground truth and the randomly sampled negative item in domain \textit{A}.
For the other domain, $\mathcal{L}_B$ can be obtained by simply replacing the notation `A' with the notation `B' in Equation~\eqref{equ:loss1} and Equation~\eqref{equ:loss2}.
\subsection{Domain-aware Profiling}
Research has proven that LLMs can be a powerful summarizer in generating users’ profiles~\cite{surveyllmesr,bridge,llmins}. However, current methods naively introduce LLMs as a profile generator, leading to rough profiles and user preferences. Thus, we propose the domain-aware profiling, containing a K-means clustering mechanism, a summarize-reform-analyze pipeline, and a preference alignment procedure.

\subsubsection{\textbf{K-means Clustering}} 
To better model the complex preferences in the mixed interaction sequence, it begins with a clustering procedure, which leverages the K-means algorithm~\cite{kmeans} to process the LLMs embeddings of each domain, \ie $\mathcal{E}_{aug}^A$ and $\mathcal{E}_{aug}^B$, and get $K$ sub-sequences of each domain based on the clusters, respectively denoted as $S^A_{g_1},S^A_{g_2},\dots,S^A_{g_C}$ and $S^B_{g_1},S^B_{g_2},\dots,S^B_{g_C}$, where $C$ is the number of cluster.

\subsubsection{\textbf{Summarize-Reform-Analyze}}\label{sec:summarize} 
After acquiring each domain's sub-sequences, we perform the summarize-reform-analyze to generate the fine-grained user profiles. In the first summarization stage, we summarize $C$ domain-specific user preferences based on the $C$ sub-sequences in domain \textit{A} and \textit{B} and denote the corresponding textual preferences as $\{P_A^c\}^C_{c=1}$ and $\{P_B^c\}^C_{c=1}$. Then, in the second reforming stage, we leverage the same domain ratio $r$ in Section~\ref{sec:intro} to balance the summaries of each domain by introducing an extra instruction to the designed prompt. Specifically, the constructed reforming instruction is a textual description of the extent of the domain imbalance issue in the interaction sequence, denoted as $\mathcal{R}$, which is first selected among 5 manually constructed options according to the value of $r$ and then reforming by powerful LLMs with the corresponding domain background. At last, in the analysis stage, we organize the $\{P_A^c\}^C_{c=1}$ and $\{P_B^c\}^C_{c=1}$ into the prompt with the pre-constructed $\mathcal{R}$ to analyze the user's cross-domain preference, denoted as $\mathcal{P}_u$.

\subsubsection{\textbf{Preference Alignment}}
To leverage the LLMs-summarized preference $\mathcal{P}_u$ in our proposed framework, we propose an alignment loss function to inject the semantic information into the user representation $\tilde{\mathbf{u}}$ obtained from the global pre-training stage. Specifically, it first converts the textual preference $\mathcal{P}_u$ to LLMs embeddings. Then, leveraging another adapter, which is similar to Equation~\eqref{equ:reducedi}, to reduce the dimension to $\mathbb{R}^d$, denoted as $\tilde{\mathbf{p}}$. Lastly, by utilizing contrastive learning~\cite{contrastivel}, we can construct one side of our proposed alignment loss $\mathcal{L}_{align}$ as follows:
\begin{equation}
   \mathcal{L}_{align}^1 = -\frac{1}{B} \sum^{B}_{i=1} \log \frac{\exp(\cos(\tilde{\mathbf{u}}_i,\tilde{\mathbf{p}}_i)/\tau)}{\sum_{j=1}^{B} \mathbb{I}_{[i\neq j]}\exp(\cos(\tilde{\mathbf{u}}_i,\tilde{\mathbf{p}}_j)/\tau)} 
\end{equation}
where $\tau$ is the temperature coefficient, $B$ is the batch size. For the other side of our designed contrastive loss $ \mathcal{L}_{align}^2$, we can simply exchange the position of $\tilde{\mathbf{u}}$ and $\tilde{\mathbf{p}}$ to obtain the value. Thus, the final alignment loss can then be formulated as $\mathcal{L}_{align} = \mathcal{L}_{align}^1 +\mathcal{L}_{align}^2$.
\subsection{Training and Inference}
In this section, we will give an overview of our training and inference process for illustration.
\subsubsection{\textbf{Training}} 
For the pre-training stage, we train a general SRS by leveraging the augmented mixed sequence $S_{aug}$. The optimization process can be formulated as:
\begin{equation}
    \arg \min_{\tilde{\theta},\Phi} \mathcal{L}_{pre} + \alpha \cdot \mathcal{L}_{align}
\end{equation}
where $\tilde{\theta}$ and $\Phi$ respectively represents the parameters of SASRec and adapter. For the fine-tuning stage, we freeze the weights of our base model. Then, leveraging two adapters to generate the domain-specific user representation. The optimization process can be formulated as follows:
\begin{equation}
  \arg \min_{\Phi_a^1,\Phi_a^2} \mathcal{L}_{A} + \mathcal{L}_{B}  
\end{equation}
where $\Phi_a^1$ and $\Phi_a^2$ respectively denote the parameters of each adapter.

\subsubsection{\textbf{Inference}} 
For inference, LLM-EDT begins with the frozen LLMs embeddings and leverages several adapters to transform them to the recommendation space. Then, we divide the prediction into two perspectives. For items in $\mathcal{A}$, it leverages both the $\mathcal{S}_{aug}$ and $\mathcal{S}_{aug}^A$ to calculate the probability shown in Equation~\eqref{equ:loss1}. For the domain \textit{B} items, we can obtain a similar probability by replacing the notation \textit{'A'} with \textit{'B'} in Equation~\eqref{equ:loss1}. Noted that, since all the augmentation procedures and aforementioned LLMs embeddings can be performed and cached in advance, we only need to load the parameters of the used adapters and SASRec when conducting the inference, which avoids introducing severe loads. 

\section{Experiment}\label{experiment}
In this section, we will present the results from comprehensive experiments to verify the effectiveness of our proposed LLM-EDT. 

% Table 1: Dataset Statistics
\begin{table}[t]
\centering
\caption{The statistics of datasets}
\resizebox{.48\textwidth}{!}{%
\begin{tabular}{c|ccc|ccc}
\toprule[1pt]
\multirow{2}{*}{\textbf{Datasets}} & 
  \multicolumn{3}{c|}{\textbf{Before Aug.}} & 
  \multicolumn{3}{c}{\textbf{After Aug.}} \\
\cmidrule(lr){2-4} \cmidrule(lr){5-7}
 & \textbf{\# Users} & \textbf{\# Items} & \textbf{Avg.len} & \textbf{\# Users} & \textbf{\# Items} & \textbf{Avg.len} \\ 
\midrule
Cloth & 9,933 & 3,278 & \multirow{2}{*}{10.71} & 9,933 & 4,204 & \multirow{2}{*}{11.04} \\
Sport & 4,263 & 1,021 & & 4,263 & 2,237 & \\
\addlinespace[0.2em]
\cmidrule[0.5pt](lr){1-7}
Electronic & 20,728 & 10,492 & \multirow{2}{*}{8.30} & 20,728 & 11,023 & \multirow{2}{*}{8.51} \\
Cell Phone & 11,762 & 2,246 & & 11,762 & 3,604 & \\
\addlinespace[0.2em]
\cmidrule[0.5pt](lr){1-7}
Food & 14,858 & 8,343 & \multirow{2}{*}{8.13} & 14,858 & 8,742 & \multirow{2}{*}{8.47} \\
Kitchen & 10,145 & 3,188 & & 10,145 & 4,378 & \\
\bottomrule[1pt]
\end{tabular}}
\label{tab:dataset}
\end{table}

% Table 2: Overall Performance (Wide)
\begin{table*}[htbp]
\centering
\tabcolsep=0.06cm 
\caption{Overall performance of LLM-EDT. The best results are bold, and the second-best are underlined. ``*'' indicates the improvements are statistically significant (i.e., one-sided t-test with \(p<0.05\) ) over baselines.}
\label{tab:performance_comparison}
\resizebox{\textwidth}{!}{%
\renewcommand{\arraystretch}{1.2}
\begin{tabular}{cc|cc|cc|cc|cc|cc|cc}
\toprule
\multicolumn{2}{c|}{\multirow{3}{*}{\textbf{Model}}} &
  \multicolumn{4}{c|}{\textbf{Amazon}} &
  \multicolumn{4}{c|}{\textbf{Amazon}} &
  \multicolumn{4}{c}{\textbf{Amazon}} \\
\cmidrule(lr){3-6} \cmidrule(lr){7-10} \cmidrule(l){11-14}
\multicolumn{2}{c|}{} &
  \multicolumn{2}{c|}{\textbf{Cloth}} &
  \multicolumn{2}{c|}{\textbf{Sport}} &
  \multicolumn{2}{c|}{\textbf{Electronic}} &
  \multicolumn{2}{c|}{\textbf{Cell Phone}} &
  \multicolumn{2}{c|}{\textbf{Food}} &
  \multicolumn{2}{c}{\textbf{Kitchen}} \\
\cmidrule(lr){3-4} \cmidrule(lr){5-6} \cmidrule(lr){7-8} \cmidrule(lr){9-10} \cmidrule(lr){11-12} \cmidrule(l){13-14}
\multicolumn{2}{c|}{} &
  \textbf{H@10} &
  \textbf{N@10} &
    \textbf{H@10} &
  \textbf{N@10} &
    \textbf{H@10} &
  \textbf{N@10} &
    \textbf{H@10} &
  \textbf{N@10} &
    \textbf{H@10} &
  \textbf{N@10} &
    \textbf{H@10} &
  \textbf{N@10} \\
\midrule
\multicolumn{1}{c|}{\multirow{3}{*}{SRS}} &
  \multicolumn{1}{c|}{GRU4Rec} & 
  0.5632 &
  0.5117 &
  0.5536 &
  0.4769 &
  0.3187 &
  0.1835 &
  0.3079 &
  0.1701 &
  0.2916 &
  0.1740 &
  0.3911 &
  0.2894 \\
\multicolumn{1}{c|}{} &
  \multicolumn{1}{c|}{BERT4Rec} & 
  0.6239 &
  0.5583 &
  0.5372 &
  0.4525 &
  0.3272 &
  0.1871 &
  0.2967 &
  0.1694 &
  0.3025 &
  0.1869 &
  0.4087 &
  0.3094 \\
\multicolumn{1}{c|}{} &
  \multicolumn{1}{c|}{SASRec} & 
  0.7081 &
  0.6499 &
  0.5911 &
  0.4872 &
  0.3501 &
  0.2194 &
  0.3212 &
  0.1979 &
  0.3250 &
  0.2078 &
  0.4211 &
  0.3208 \\ 
\midrule
\multicolumn{1}{c|}{\multirow{4}{*}{CDSR}} & % Changed from 3 to 4
  \multicolumn{1}{c|}{C2DSR} & 
  0.7166 &
  0.6701 &
  0.5948 &
  0.5627 &
  0.4559 &
  0.3156 &
  0.2819 &
  0.1789 &
  0.3416 &
  0.2293 &
  0.4424 &
  0.3417 \\
\multicolumn{1}{c|}{} &
  \multicolumn{1}{c|}{TriCDR} & 
  0.7219 &
  0.6760 &
  0.6114 &
  0.5703 &
  0.4532 &
  0.3196 &
  0.2616 &
  0.1575 &
  0.3457 &
  0.2301 &
  0.4446 &
  0.3442 \\
\multicolumn{1}{c|}{} &
  \multicolumn{1}{c|}{SyNCRec} & 
  0.7301 &
  0.6783 &
  0.6172 &
  0.5594 &
  0.4272 &
  0.3014 &
  0.3332 &
  0.2000 &
  0.3362 &
  0.2116 &
  0.4383 &
  0.3211 \\ 
\multicolumn{1}{c|}{} &
  \multicolumn{1}{c|}{ABXI} & 
  0.7474 &
  0.6972 &
  0.6440 &
  0.6083 &
  0.4623 &
  0.3047 &
  0.3282 &
  0.2153 &
  0.3652 &
  0.2453 &
  0.4682 &
  0.3546 \\
\midrule
\multicolumn{1}{c|}{\multirow{3}{*}{\centering LLMESR}} & 
  \multicolumn{1}{c|}{LLM-ESR} & 
  0.7433 &
  0.6804 &
  0.6394 &
  0.5911 &
  0.4548 &
  0.2767 &
  0.3411 &
  0.2103 &
  0.3667 &
  0.2487 &
  0.4551 &
  0.3485 \\
\multicolumn{1}{c|}{} &
  \multicolumn{1}{c|}{LLMCDSR} & 
  0.7646 &
  0.7023 &
  {\underline {0.7138}} &
  \underline{ 0.6379} &
  \underline{ 0.4900} &
  \underline{ 0.3292} &
  \underline{0.3586} &
  \underline{ 0.2417} &
  0.3746 &
  0.2594 &
  0.4779 &
  0.3601 \\
\multicolumn{1}{c|}{} &
  \multicolumn{1}{c|}{LLM4CDSR} & 
  \underline{ 0.7882} &
  \underline{ 0.7205} &
  0.7021 &
  0.6249 &
  0.4818 &
  0.3277 &
  0.3423 &
  0.2229 &
  \underline{ 0.3976} &
  \underline{ 0.2702} &
  \underline{ 0.4901} &
  \underline{ 0.3768} \\ 
\midrule
\multicolumn{1}{c|}{\multirow{2}{*}{Ours}} &
  \multicolumn{1}{c|}{LLM-EDT} & 
  \textbf{0.8104}\(^*\) &
  \textbf{0.7488}\(^*\) &
  \textbf{0.7649}\(^*\) &
  \textbf{0.6848}\(^*\) &
  \textbf{0.5289}\(^*\) &
  \textbf{0.3775}\(^*\) &
  \textbf{0.4553}\(^*\) &
  \textbf{0.3041}\(^*\) &
  \textbf{0.4349}\(^*\) &
  \textbf{0.3109}\(^*\) &
  \textbf{0.5406}\(^*\) &
  \textbf{0.4236}\(^*\) \\
\multicolumn{1}{c|}{} &
  \multicolumn{1}{c|}{Improv. (\%)} & 
  2.81 &
  3.93 &
  7.16 &
  7.35 &
  7.94 &
  14.67 &
  26.96 &
  25.82 &
  9.38 &
  15.06 &
  10.30 &
  12.42
\\
\bottomrule
\end{tabular}}
\end{table*}

\subsection{Experiment Settings}
\subsubsection{\textbf{Dataset}}
We conduct our experiments with three public datasets, \ie \textbf{Cloth - Sport}, \textbf{Electronic - Cell Phone}, and \textbf{Food - Kitchen}, which originate from the sub-categories of the Amazon dataset\footnote{\url{https://cseweb.ucsd.edu/jmcauley/datasets.html\#amazon_reviews}}. For details, we summarize the statistics of the selected datasets before and after our augmentation as presented in Table~\ref{tab:dataset}.

\subsubsection{\textbf{Baselines}}
To validate the superiority of LLM-EDT, we conduct our comparison experiment with various up-to-date baselines, which can be categorized into three groups. 
\begin{itemize}[leftmargin=*]
    \item \textbf{{Sequential Recommender Systems (SRS)}}: GRU4Rec~\cite{GRU4Rec},
     BERT4Rec~\cite{Bert4rec}, and SASRec~\cite{SASRec}.
    \item \textbf{{Cross-domain Sequential Recommendation (CDSR)}}: C2DSR~\cite{CDSR1}, TriCDR~\cite{TriCDR}, SyNCRec~\cite{SyNCRec}, and ABXI~\cite{ABXI}.
    \item \textbf{{LLMs-enhanced Sequential Recommendation (LLMESR)}}: LLM-ESR~\cite{LLM-ESR}, LLMCDSR~\cite{llmcdsr}, and LLM4CDSR~\cite{bridge}.
\end{itemize}

\subsubsection{\textbf{Implementation Details}}
In this section, we detail the implementation of our LLM-EDT. The code is released online\footnote{\url{https://anonymous.4open.science/r/LLM-EDT-Annoy}}. For the hardware selection, we conduct all of our experiments on a single NVIDIA 4090 GPU. The other necessary appliances include Python 3.11, PyTorch 2.4.1, etc. In the {Transferable Item Augmenter}, the number of clusters $K$ is set to 10. We conduct both the embedding and augmentation processes by leveraging the API of DeepSeek\footnote{\url{https://api.deepseek.com}}. Regarding the architecture setting, we set the hidden size to 128 and batch size to 512 for our LLM-EDT and all the compared baselines for fairness. For the training procedure, we set the batch size as 512 and the learning rate as 0.001 for all datasets in the {Global Pre-training stage}, and set the adapter size to 64 with a smaller learning rate, equal to 1e-5, in the {Domain Fine-tuning Stage}. For the weight of the alignment loss $\mathcal{L}_{align}$, we search the value $\alpha$ from $\{0.1,0.3,0.5,0.7,1.0\}$ for all the datasets. All the presented results are the average values of three results with random seeds $\{42,44,46\}$. Other implementation details are the same as the previous work~\cite{bridge}.

\subsubsection{\textbf{Evaluation Metrics}}
To provide a comprehensive view of our framework, we adopt the \textit{Hit Rate} and \textit{Normalized Discounted Cumulative Gain}, all truncated at 10, denoted as $\mathbf{H@10}$ and $\mathbf{N@10}$.

\subsection{Overall Performance} %(RQ1)}
In this section, we compare our proposed method with various baselines across three public datasets. As shown in Table~\ref{tab:performance_comparison}, the proposed LLM-EDT demonstrates remarkable performance with significant improvements ranging from $2.81\%$ to $26.96\%$ over the best baselines, highlighting the effectiveness of our designed structure. 

For a more detailed analysis, we first observe that general SRS models trail behind other baselines (\ie CDSR and LLM-ESR) by a significant margin. This is primarily attributed to data sparsity and the loss of semantic information inherent in single-domain sequences. Regarding conventional CDSR baselines, models such as C2DSR, TriCDR, and SyNCRec exhibit better performance than SRS. This is due to their capability to capture both domain-specific and domain-shared features, which is particularly beneficial for predicting non-dominant domains (\eg Sport and Kitchen). Notably, among this category, ABXI achieves the most competitive performance, consistently outperforming other traditional CDSR methods across most metrics. This indicates that ABXI effectively mitigates the negative transfer problem by leveraging its task-guided prediction. However, despite these gains, ABXI and other conventional CDSR models still rely heavily on ID-based patterns, falling short in effectively leveraging the rich semantic information. 

Benefiting from the reasoning and representation capabilities of Large Language Models (LLMs), LLM-enhanced SRS has emerged as a powerful technique that leverages both ID-based and semantic information. As a representative work, LLM-ESR demonstrates impressive performance by combining collaborative and semantic views. Nevertheless, because it focuses on single-domain prediction, its performance on CDSR tasks is lower than that of LLM4CDSR and LLMCDSR. Among LLM-based CDSR methods, LLMCDSR achieves remarkable results, especially in non-dominant domains, thanks to its latent item generation and meta-learning network. However, it suffers from irrelevant noise introduced by naive augmentation and lacks fine-grained representation for domain-specific features, leading to sub-optimal performance. Similarly, LLM4CDSR employs a three-thread structure to enhance user and item representations via LLMs. While strong, it fails to fully capture the correlations among these threads and overlooks the noise introduced by coarse-grained user profiles, resulting in sub-optimal performance.

\subsection{Ablation Study}% (RQ2)}
To validate the effectiveness, we conduct the ablation experiment in this section. Specifically, we categorize the variants as: 
\begin{itemize}[leftmargin=*]
    \item \textbf{\textit{w/o} Aug.}: This variant removes the whole data augmentation and leverage the raw sequence to perform the next-item prediction. 
    \item \textbf{\textit{w/o} Filtering}: This variant removes the noise filtering process and inserts the initially generated samples into the raw sequence.
    \item \textbf{\textit{w/o} Inserting}: This variant removes the inserting process and directly arranges the generated sample at the end of the sequence. 
    \item \textbf{Random Inserting}: This variant randomly inserts the generated sample in the mixed interaction sequence. 
    \item \textbf{\textit{w/o} DFT}: This variant removes the domain fine-tuning process to validate the effectiveness of {Dual Phase Training}. 
    \item \textbf{\textit{w/o} Profiling}: This variant removes the {Domain-aware profiling} module by eliminating $\mathcal{L}_{align}$.
    \item \textbf{\textit{w/o} S-R-A}: This variant removes the summarize-reform-analyze pipeline when generating users' profiles and replaces it with the naive LLMs summarization.
\end{itemize}

% Table 3: Ablation Results
\begin{table}[t]
\centering
\caption{Ablation study results of LLM-EDT}
\label{tab:ablation_results}  
%\resizebox{.48\textwidth}{!}{
\begin{tabular}{@{}c c c c c@{}}
\toprule
\multirow{2}{*}{\textbf{Variants}} & \multicolumn{2}{c}{\textbf{Cloth}} & \multicolumn{2}{c}{\textbf{Sport}} \\
\cmidrule(lr){2-3} \cmidrule(l){4-5}
 & \textbf{H@10 }& \textbf{N@10} & \textbf{H@10} & \textbf{N@10} \\ 
\midrule
\textbf{LLM-EDT}        & \textbf{0.8078}  & \textbf{0.7432}  & \textbf{0.7529}  & \textbf{0.6779}  \\ 
\textit{w/o} Aug.       & 0.7919  & 0.7311  & 0.6854  & 0.6291  \\
\textit{w/o} Filtering.       & 0.7724  & 0.7148  & 0.6670  & 0.6079  \\
\textit{w/o} Inserting.       & 0.7802  & 0.7195  & 0.6703  & 0.6214  \\
Random Inserting. & 0.7842  & 0.7235  & 0.6774  & 0.6377 \\
\textit{w/o} DFT        & 0.7581 & 0.6753 & 0.7049 & 0.6022 \\
\textit{w/o} Profiling & 0.7951 & 0.7377  & 0.7399 & 0.6617  \\
\textit{w/o} S-R-A        & 0.7901 & 0.7321 & 0.7278 & 0.6526 \\
\bottomrule
\end{tabular}
%}
\end{table}

We test these variants on Cloth - Sport, and present the results in Table~\ref{tab:ablation_results}. According to the results, we can draw our conclusion: %to answer \textbf{RQ2}:
\begin{itemize}[leftmargin=*]
    \item Our proposed {transferable item augmenter} with its noise filtering and domain-aware insertion modules contributes to improving the recommendation performance for all domains.
    \item {Dual Phase training} enhances the LLM-EDT framework to capture fine-grained representations for individual domains, thereby improving performance.
    \item {Domain-aware profiling} significantly improves the performance on non-dominant domains compared to rough profiling~\cite{bridge}, which helps to alleviate the domain imbalance issue in modeling user global preference.
\end{itemize}

\begin{figure}[!t]
    \centering
    \includegraphics[width = \linewidth]{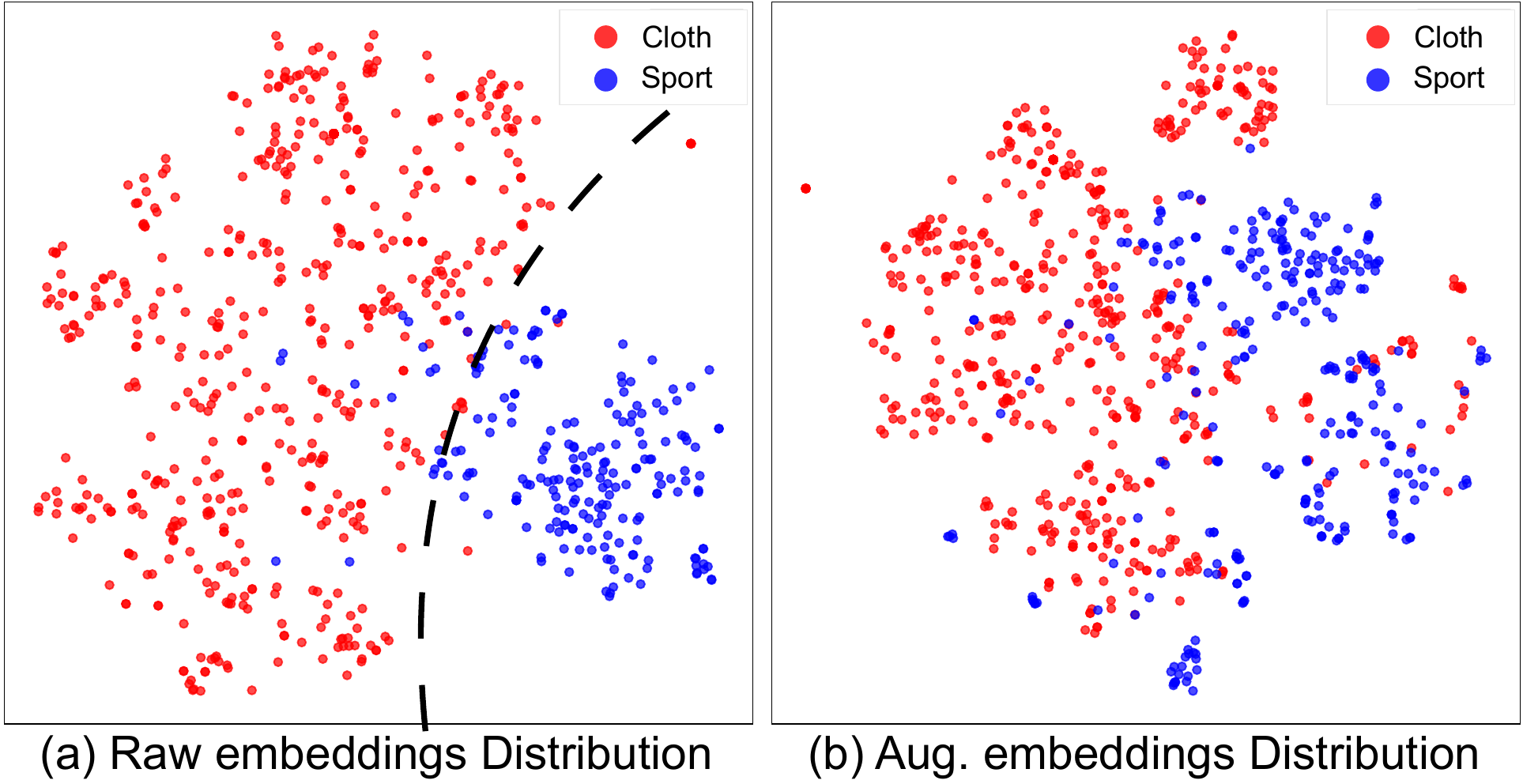}
    \caption{Distribution Comparison in Cloth - Sport.}
        \label{fig:dis_clo}
\end{figure}

\subsection{Transferability Analysis} \label{sec:trans}
In this section, we present how we enhance the transferability by leveraging the designed {Transferable Item Generator}. As previously discussed, transferable items can be defined as items that belong to one domain but have relatively high similarity with another domain. We present the item embeddings distribution processed by t-SNE in Figure~\ref{fig:dis_clo}.  
Specifically, observing from the results in Figure~\ref{fig:dis_clo} (a), there is a clear margin between the two domains (respectively using red and blue marks to represent) before augmentation in semantic space, presenting the need for transferable items to act as a bridge. Compared to the original item embeddings, the augmented item embeddings, presented in Figure~\ref{fig:dis_clo} (b), show
a more intensive distribution that embeddings from each domain present closer distances, which means we have generated the transferable items and constructed an effective bridge between domains. The model performance of variant \textit{w/o} Aug., shown in the above Table~\ref{tab:ablation_results}, further validates the transferability it brings.

\subsection{Cold-Start Analysis} %(RQ4)}
\begin{figure}[!t]
    \centering
    \includegraphics[width = \linewidth]{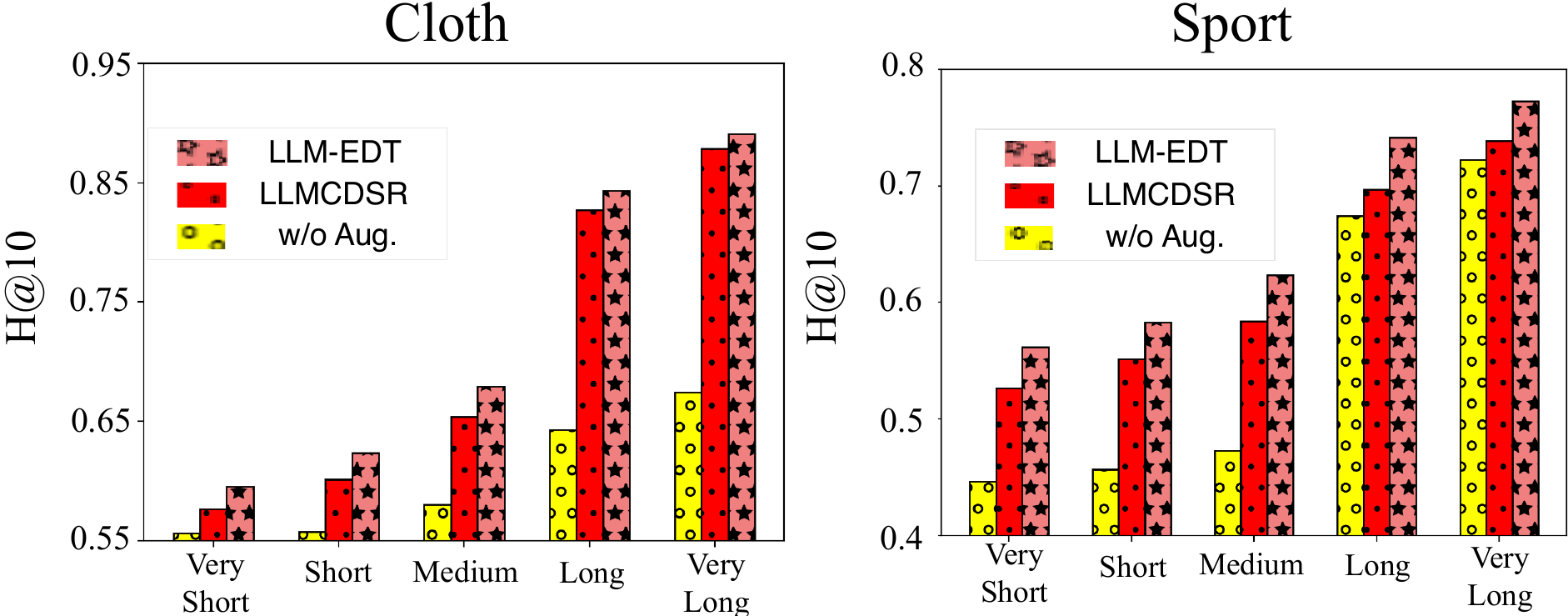}
    \caption{Performance on grouped users in Cloth-Sports}
        \label{fig:cold_A}
\end{figure}

Noted that, since our proposed {transferable item augmenter} aims to generate items that act as a bridge to transfer information between domains, it can not only improve the recommendation performance on each domain, but also help to alleviate the cold-start problem in CDSR originating from the imbalanced domain problem. Specifically, we categorize the Cloth - Sport dataset into five subsets strictly based on user interaction length: "Very Short" (0-5), "Short" (5-10), "Medium" (10-15), "Long" (15-20), "Very Long" (20, inf). Then, we compare our proposed LLM-EDT with the variant \textit{w/o} Aug, and LLMCDSR, which also leverages data augmentation, to show our superiority. The experimental results are presented in Figure~\ref{fig:cold_A}. We can observe that both augmentation methods outperform the one that removes the augmentation module, verifying the effectiveness of data augmentation in alleviating the cold-start problem in CDSR. Moreover, observing the results between LLM-EDT and LLMCDSR from both the Cloth perspective and the Sport perspective, LLM-EDT outperforms LLMCDSR in all user lengths, which further validates that our designed structure can generate more effective items to alleviate the cold-start problem in CDSR.

% ---------------------------------------------------------------------
% IMPORTANT: The EFFICIENCY TABLE code is moved here (before Generality Validation)
% to allow the table* to float to the top of the next page (Efficiency section).
% ---------------------------------------------------------------------

% ---------------------------------------------------------------------

\subsection{Hyperparameter Analysis}% (RQ5)}
\begin{figure}[H]
    \centering
    \includegraphics[width = \linewidth]{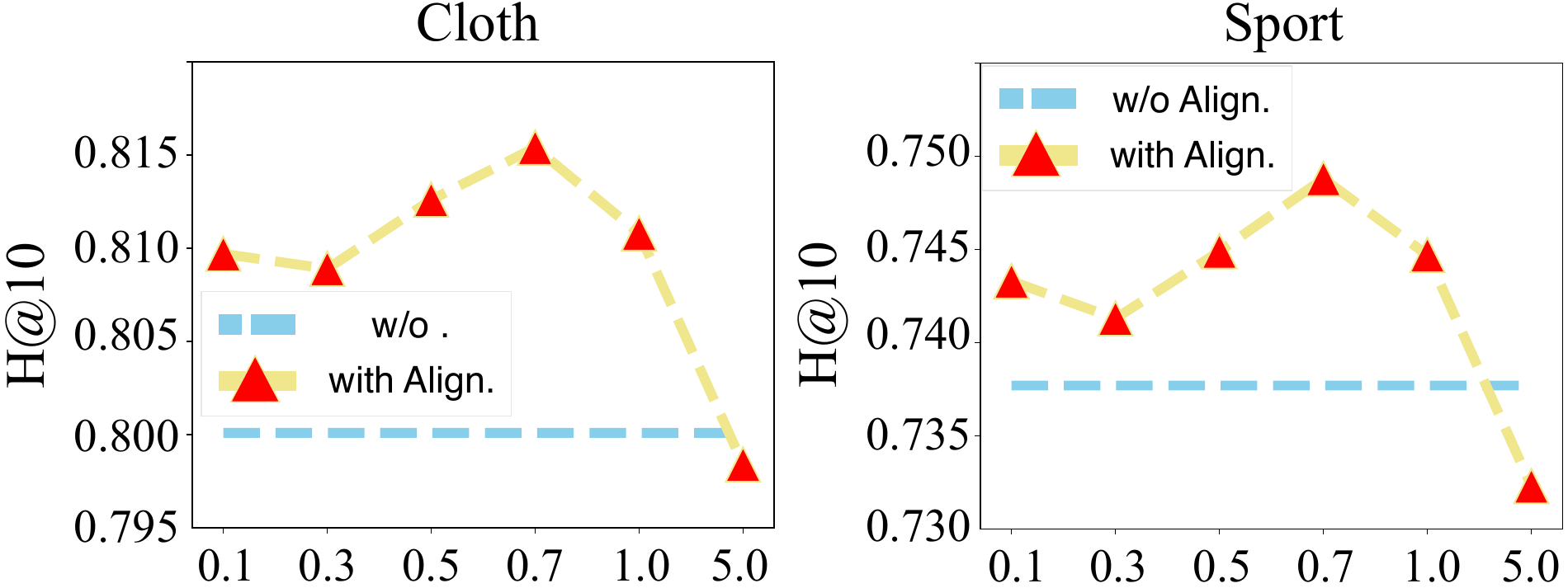}
    \caption{Hyper-parameter Results on Amazon Cloth-Sport.}
        \label{fig:hyper}
\end{figure}

To perform our hyper-parameter experiment, we vary the weights of the alignment loss $\mathcal{L}_{align}$ and present the performance trends conducted on Cloth - Sport in Figure~\ref{fig:hyper}. We can observe that the H@10 shows a rising-descending trend within $\left[0.1,5\right]$ for the Cloth aspect, and the best value is 0.7 broadly. For the Sport aspect, it presents a similar pattern but a higher best value, equal to 1.0 broadly. The reason can be concluded as the relatively less information compared to Cloth, which needs to be further aligned.

Furthermore, we also conduct the performance of our LLM-EDT with different K-Means Cluster $k$ and different threshold values $\tau$ to validate the effectiveness and draw our conclusion:
\begin{itemize}[leftmargin=*]
    \item For the similarity threshold in Eqution~\ref{equ:tau}, extremely high values (\eg 0.99) are too strict and yield few qualified items. Performance under different thresholds in Table~\ref{tab:tau} supports our choice.
    
    \item K-Means adaptively selects representative items to minimize distribution shift. Cluster count mainly affects categorization and does not significantly impact final performance, as shown in the Table~\ref{tab:cluster}. 
\end{itemize}

\label{sec:hyper}

% Merged Table: Tau and Cluster side-by-side
\begin{table}[t]
    \centering
    \begin{minipage}{0.48\linewidth}
        \centering
        \caption{Comparison with $\tau$}
        \label{tab:tau}
        %\resizebox{\linewidth}{!}{
            \begin{tabular}{lcc}
                \toprule
                \textbf{Model} & \textbf{Cloth} & \textbf{Sport} \\
                \midrule
                $\tau=0.85$ & 0.8078 & 0.7573 \\
                $\tau=0.90$ & \textbf{0.8104} & \textbf{0.7649} \\
                $\tau=0.95$ & 0.8066 & 0.7587 \\
                \bottomrule
            \end{tabular}
        %}
    \end{minipage}
    \hfill
    \begin{minipage}{0.48\linewidth}
        \centering
        \caption{Comparison with $k$}
        \label{tab:cluster}
        %\resizebox{\linewidth}{!}{
            \begin{tabular}{lcc}
                \toprule
                \textbf{Model} & \textbf{Cloth} & \textbf{Sports} \\
                \midrule
                $k=5$ & 0.8099 & 0.7487 \\
                $k=10$ & \textbf{0.8104} & \textbf{0.7649} \\
                $k=15$ & 0.7924 & 0.7601 \\
                \bottomrule
            \end{tabular}
        %}
    \end{minipage}
\end{table}

\subsection{Transferable Augmenter Analysis}
To validate the performance of our Augmenter, we conduct experiments in two dimensions. 
First, we computed cosine similarity between original and augmented user profiles in Amazon Cloth-Sport dataset. Results in Table~\ref{tab:injection} show that high similarity is maintained across different injection ratios, validating that our augmentation does not significantly affect the user's original preference, thereby showing our generation quality. Next, since our item augmenter may affect the pre-defined domain ratio $r$ in section~\ref{sec:intro}, we gradually inject an increasing number of enhanced items in Table~\ref{tab:domain_ratio}, which demonstrates that the augmentation preserves the domain distribution and improves the performance stably.

\begin{table}[H]
    \centering
    \caption{Similarity comparison for different injection ratios}
    \label{tab:injection}
    \resizebox{.8\linewidth}{!}{
        \begin{tabular}{lccc}
            \toprule
            \textbf{Inject Ratio} & \textbf{30\%} & \textbf{60\%} & \textbf{100\%} \\
            \midrule
            Similarity & 98.76\% & 98.28\% & 97.91\% \\
            \bottomrule
        \end{tabular}
    }
\end{table}

\begin{table}[H]
    \centering
    \caption{Performance under different domain ratios}
    \label{tab:domain_ratio}
    %\resizebox{\linewidth}{!}{
        \begin{tabular}{lccc}
            \toprule
            \textbf{Model} & \textbf{Condition} & \textbf{Cloth} & \textbf{Sports} \\
            \midrule
            \multirow{3}{*}{LLM-EDT} & $r \ge 0.95$ & 0.7945 & 0.7421 \\
            & $0.95 \ge r \ge 0.9$ & 0.8035 & 0.7589 \\
            & $r < 0.9$ & 0.8104 & 0.7649 \\
            \bottomrule
        \end{tabular}
    %}
\end{table}

\subsection{Generality Validation}% (RQ6)}
We conduct our generality validation in two aspects. 
To validate the generality of our LLM-agnostic framework, we compare the performance of two different variants, \ie replacing our used LLM from DeepSeek to GPT-3.5 and text-embedding-002. As shown in the Table~\ref{tab:ge}, LLM-EDT with different LLMs yields stable performance, supporting the use of cost-effective or open-source LLMs.
%\begin{table*}[t]
%\centering
%\caption{Efficiency comparison on Cloth - Sport dataset.}
%\label{tab:efficiency_comparison}
%\resizebox{.48\textwidth}{!}{
%\begin{tabular}{@{}{c} l *{5}{c} @{}{c}}
%\toprule
%\textbf{Models} & \textbf{SASRec} & \textbf{C2DSR} & \textbf{LLMCDSR} & \textbf{LLM4CDSR} & \textbf{LLM-EDT} \\
%\midrule
%Parameter (\# M)   & \textbf{3.99}    & 7.81    & 6.38    & 7.72    & \underline{4.12} \\
%Inference Time (\# s) & 7.16    & 5.01    & 5.41    & \textbf{4.7} & \underline{4.92} \\
%\bottomrule
%\end{tabular}
%}
%\end{table*}
\begin{table}[t]
    \centering
    \caption{Performance comparison with different LLMs}
    \label{tab:ge}
    \resizebox{\linewidth}{!}{
        \begin{tabular}{lcccc}
            \toprule
            \multirow{2}{*}{\textbf{Variants}} & \multicolumn{2}{c}{\textbf{Cloth}} & \multicolumn{2}{c}{\textbf{Sport}} \\
            \cmidrule(lr){2-3} \cmidrule(lr){4-5}
             & \textbf{H@10} & \textbf{N@10} & \textbf{H@10} & \textbf{N@10} \\ % 注意这里第一个单元格留空
            \midrule
            With DeepSeek & 0.8104 & 0.7488 & 0.7649 & 0.6848 \\
            With GPT3.5   & 0.8091 & 0.7452 & 0.7581 & 0.6792 \\
            \bottomrule
        \end{tabular}
    }
\end{table}
To validate the generality of our model-agnostic framework, we compare the most competitive baseline, \ie LLM4CDSR, with LLM-EDT by replacing the backbone with two traditional models, \ie GRU4Rec and BERT4Rec. The presented results in Table~\ref{tab:gv} further validate that our LLM-EDT can achieve the state-of-the-art performance with different kinds of backbones, indicating the potential application of LLM-EDT in real-world scenarios.

\begin{table}[t]
\tabcolsep=0.05cm 
\centering
\caption{Model Performance on different backbones.}
\label{tab:gv}
\resizebox{.48\textwidth}{!}{
\begin{tabular}{@{}c *{8}{c} @{}}
\toprule
\multirow{3}{*}{\textbf{Model}} & 
\multicolumn{4}{c}{\textbf{GRU4Rec}} & 
\multicolumn{4}{c}{\textbf{BERT4Rec}} \\
\cmidrule(lr){2-5} \cmidrule(l){6-9}
& \multicolumn{2}{c}{\textbf{Cloth}} & \multicolumn{2}{c}{\textbf{Sport}} & 
\multicolumn{2}{c}{\textbf{Cloth}} & \multicolumn{2}{c}{\textbf{Sport}} \\
\cmidrule(lr){2-3} \cmidrule(lr){4-5} \cmidrule(lr){6-7} \cmidrule(l){8-9}
& \textbf{H@10} & \textbf{N@10} & \textbf{H@10} & \textbf{N@10} & \textbf{H@10} & \textbf{N@10} & \textbf{H@10} & \textbf{N@10} \\
\midrule
- & 0.5632 & 0.5117 & 0.5536 & 0.4769 & 0.6239 & 0.5583 & 0.5372 & 0.4525 \\
LLMCDSR & 0.7703 & 0.6865 & 0.6810 & 0.6049 & 0.7835 & 0.7084 & 0.6826 & 0.6088 \\
LLM4CDSR & 0.7599 & 0.6672 & 0.7055 & 0.6201 & 0.7713 & 0.6908 & 0.7035 & 0.6302 \\
LLM-EDT & \textbf{0.8104} & \textbf{0.7226} & \textbf{0.7573} & \textbf{0.6546} & \textbf{0.8113} & \textbf{0.7409} & \textbf{0.7814} & \textbf{0.6865} \\
\bottomrule
\end{tabular}
}
\end{table}

\subsection{Efficiency Analysis}

%\begin{table}[t]

%    \centering
%    \caption{Cost Per User using DeepSeek}
%    \label{tab:price}
    %\resizebox{\linewidth}{!}{
%        \begin{tabular}{ccc}
%            \toprule
%            Embedding & Generation & Summarization \\
%            \midrule
%            \$0.000009 & \$0.000194 & \$0.000477 \\
%            \bottomrule
%        \end{tabular}
    %}

%\end{table}

\begin{figure}[H]
    \centering
    \includegraphics[width = \linewidth]{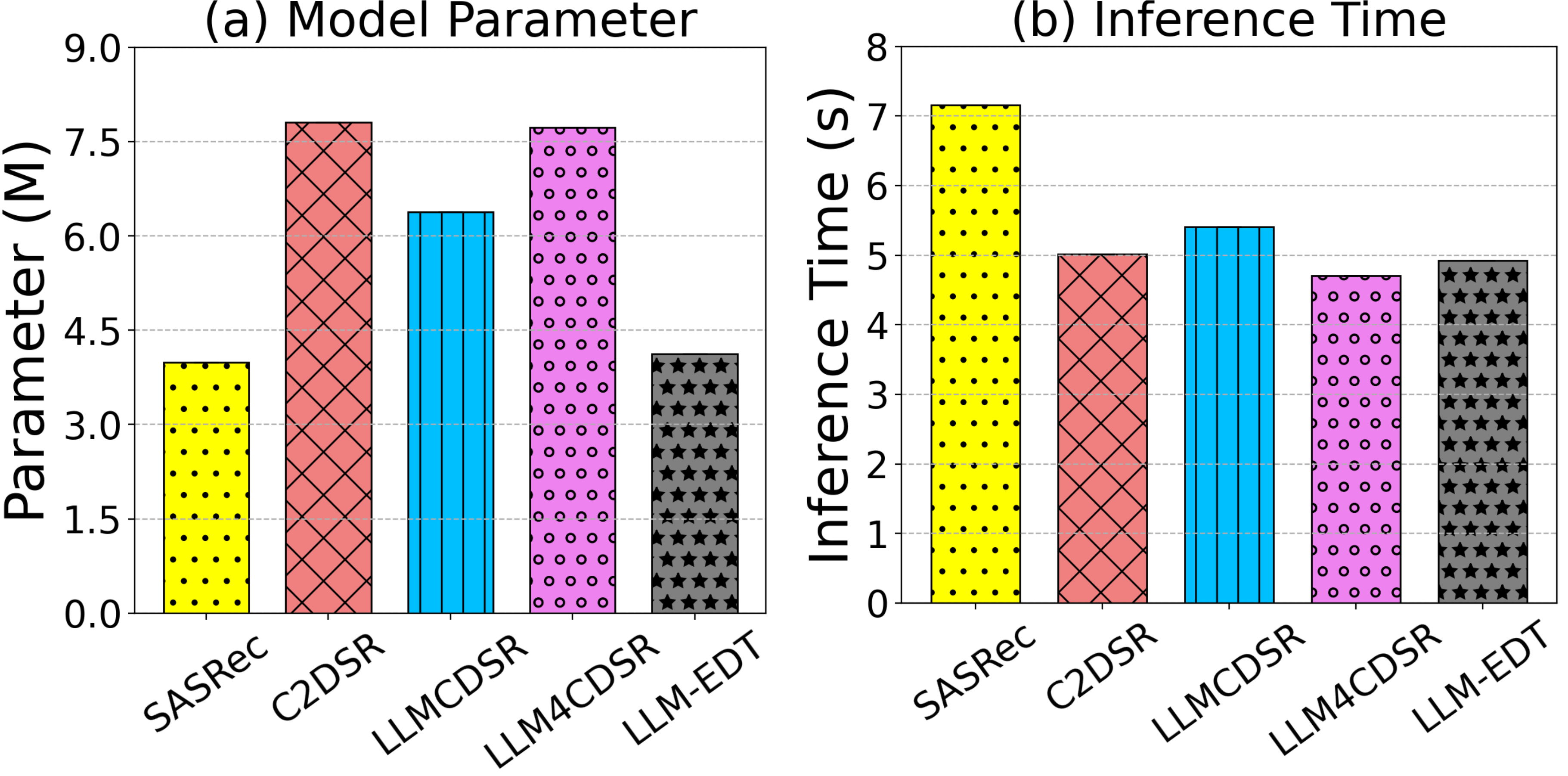}
    \caption{Model Parameters and Inference Time Comparison among different baselines.}
        \label{fig:effi}
\end{figure}

To further validate the efficiency of our proposed LLM-EDT, we first test the API cost per user. The embedding stage is very affordable, costing only $0.00009 \$$. In the generation and summarization stages, the API costs are $0.000194\$$ and $0.000477 \$$ per user, validating the scalability of our framework. Moreover, we conduct the efficiency experiment on three datasets compared with the current CDSR models shown in Figure~\ref{fig:effi}. We can then draw our conclusion:
\begin{itemize}[leftmargin=*]
    \item The parameter size of LLM-EDT is only marginally larger than the lightweight SASRec baseline, while much smaller than current CDSR frameworks.
    \item The inference latency of LLM-EDT is slightly higher than that of LLM4CDSR due to its two domain-adaptation adapters, but it achieves state-of-the-art (SOTA) recommendation performance while maintaining near-SOTA inference time and second-best parameter efficiency among comparative models.
    \item The API cost per user for different generation stages is decent and acceptable, validating the scalability of the proposed LLM-EDT.
\end{itemize}

\section{Related Work}
\subsection{Cross-domain Sequential Recommendation} 
To address the inherent limitations of Single-domain Sequential Recommender Systems (SRS), such as data sparsity~\cite{sparsity}, CDSR leverages interaction records from multiple domains to enrich user representations~\cite{cdsrsurvey1}. Beyond classic CDSR settings, recent multi-domain recommendation studies also improve knowledge sharing across domains by introducing hyper-adapters or mixture-of-experts mechanisms~\cite{li2023hamur,zhang2024m3oe}. However, the development of CDSR still faces significant hurdles, primarily the \textbf{Domain Imbalance} and \textbf{Domain Transition} challenges. Early works addressed domain imbalance via transfer learning~\cite{transfer1,transfer2}. Representative examples include Gromov-Wasserstein guided cross-domain representation learning~\cite{li2022gromov} and instance transfer methods such as AutoTransfer~\cite{gao2023autotransfer}, which strengthen information flow across domains.

To better capture complex interaction patterns, GNN-based approaches like C2DSR~\cite{CDSR1} have been proposed. Unlike prior pipeline methods, C2DSR jointly learns single-domain and cross-domain representations by modeling both intra-sequence item relationships and inter-sequence collaborative signals via a graph neural network. Other studies further improve CDSR by combining internal multi-interest exploration with external domain alignment~\cite{CDSR2}. Regarding domain transition, recent studies focus on balancing domain-shared and domain-specific features~\cite{cdsrsurvey2}. For instance, TriCDR~\cite{TriCDR} employs a triple-sequence attention network with contrastive learning to align source, target, and mixed sequences, capturing global interests across domains while retaining domain-specific sensitivity.

Despite these advances, negative transfer remains a critical issue. SyNCRec~\cite{SyNCRec} explicitly tackles this by employing an asymmetric cooperative network that estimates the negative transfer impact per domain to adaptively weight prediction losses. Similarly, ABXI~\cite{ABXI} identifies the prediction mismatch in timestamp-based alignment and introduces a task-guided alignment mechanism to extract domain-invariant interests. Nevertheless, these approaches often overlook fine-grained domain-specific details during the transfer process. In contrast, our work tackles both issues via a transferable item augmenter and dual-phase training.

\subsection{LLM-enhanced Sequential Recommendation}
LLMs enhance SRS through semantic understanding and reasoning~\cite{surveyllmesr,llmsurvey}. Recent studies further revisit LLM architectures for sequential recommendation and summarize the methodological landscape of LLM-enhanced recommender systems~\cite{wang2025rethinking,liu2025large}. Common strategies use LLMs as encoders (\eg SAID~\cite{SAID}, LLMEmb~\cite{LLMemb}). LLMs have also been used to improve long-tailed sequential recommendation and model long textual user behaviors~\cite{LLM-ESR,lengthy}. Meanwhile, in broader recommendation settings, LLMs have also been used for prompt-enhanced multi-scenario recommendation, LLM-enhanced multi-scenario modeling, and unified multi-domain CTR prediction~\cite{wang2023plate,wang2024llm4msr,fu2025unified}. 
In the context of CDSR, LLMCDSR~\cite{llmcdsr} utilizes LLMs to generate "pseudo items" that bridge non-overlapped users. It employs a Relevance-Aware Meta Recall Network (RMRN) to filter these generations; however, it still risks the \textbf{irrelevant noise} problem where hallucinatory content degrades recommendation accuracy. Another method, LLM4CDSR~\cite{bridge}, follows a tri-thread framework and leverages hierarchical LLM profiling to unify semantic item representations across domains. Related personalization studies also show that LLMs can support user modeling through adaptive retrieval and memory recollection mechanisms~\cite{zhang2026evoking, zhang2026memsearcho1empoweringlargelanguage}. While existing CDSR-oriented methods attempt to balance the domain gap through contrastive tasks or profile summarization, they often fail to acquire fine-grained user preferences by relying on naive profiling or coarse cross-domain transfer, leading to the \textbf{rough profiling} problem. In contrast, our proposed LLM-EDT explicitly tackles these issues by employing the transferable item augmenter and domain-aware profiling module.

\section{Conclusion}
In this paper, we identify the limitations of current CDSR methods and propose a novel framework, LLM-EDT. Specifically, to alleviate the domain imbalance issue while introducing less noise, we propose the transferable item augmenter to generate items for transferring information across domains. Then, we introduce a dual-phase training framework to empower the domain-specific thread with the domain-shared background, thereby mitigating the domain transition issue. At last, a domain-aware profiling module is designed to generate fine-grained user profiles and align them with the domain-shared user representation. Comprehensive experiments on three public datasets validate the effectiveness of our proposed LLM-EDT framework. 

\begin{acks}
    This research was partially supported by National Natural Science Foundation of China (No.62502404), Hong Kong Research Grants Council (Research Impact Fund No.R1015-23, Collaborative Research Fund No.C1043-24GF, General Research Fund No. 11218325), Institute of Digital Medicine of City University of Hong Kong (No.9229503), Huawei (Huawei Innovation Research Program), Tencent (Tencent Rhino-Bird Focused Research Program, Tencent University Cooperation Project), Didi (CCF-Didi Gaia Scholars Research Fund), Kuaishou (CCF-Kuaishou Large Model Explorer Fund No. 2025008, Kuaishou University Cooperation Project), and Bytedance
\end{acks}
\bibliographystyle{ACM-Reference-Format}
\balance
\bibliography{Reference}
\clearpage

\end{document}